\documentclass[english]{svjour3}
\usepackage[T1]{fontenc}
\usepackage[latin9]{inputenc}
\usepackage{url}
\usepackage{amsmath}
\usepackage{amssymb}
\usepackage{esint}

\makeatletter

\providecommand{\tabularnewline}{\\}

\RequirePackage{fix-cm}

\smartqed  % flush right qed marks, e.g. at end of proof

\makeatother

\usepackage{babel}
\begin{document}
\title{Annular structures in perturbed low mass disc-shaped gaseous nebulae
II : general and polytropic models}
\titlerunning{Annular structures in gaseous nebulae II}
\author{Vladimir Pletser }
\authorrunning{V. Pletser}
\institute{lnstitut d'Astronomie et de Geophysique G.Lemaitre, Catholic University
of Louvain, Louvain-la-Neuve, Belgium \\
\emph{Present address:} Blue Abyss, Newquay, Cornwall, United Kingdom,
Vladimir.Pletser@blueabyss.uk}
\institute{ORCID 0000-0003-4884-3827}
\date{Received: date / Accepted: date}
\maketitle
\begin{abstract}
This is the second of two papers where we study additional analytical
solutions of a bidimensional low mass gaseous disc rotating around
a central mass and submitted to small radial perturbations. In a first
Paper, hydrodynamics equations were solved for the equilibrium and
perturbed configurations and a wave-like equation for the gas perturbed
specific mass was deduced and solved analytically for several cases
of exponents of the power law distributions of the unperturbed specific
mass and sound speed. In this paper, two other general cases of exponents,
including a polytropic case, are solved analytically for small frequencies
of the perturbations. Similar conclusions to the ones of Paper I are
found, namely that the maxima of the gas perturbed specific mass are
exponentially spaced and that their distance ratio is a constant,
function of disc characteristics and of the perturbations frequency.
Gaseous annular structures would eventually be formed in the disc
by inward and outward gas flows from zones of minima toward zones
of maxima of perturbed specific mass.

\end{abstract}

\section{Introduction\label{sec:Introduction}}

In a previous paper (\cite{Paper1} hereafter referred to as Paper
1), we presented analy1ical solutions of perturbations propagating
in a differentially rotating, axisymmetric, thin gaseous nebular disc,
undergoing poly1ropic transformations of index $\gamma$. Viscous
and magnetohydrodynamic forces were neglected in the disc. The specific
mass and sound speed in the disc at equilibrium had power law distributions
in the radial distance $r$, respectively $\rho\sim r^{d}$ and $c\sim r^{\frac{s}{2}}$.
Searching for solutions yielding annular structures to appear in the
disc when submitted to small radial periodic perturbations, hydrodynamics
equations were solved analy1ically for the equilibrium and perturbed
configurations, for two particular cases ($d=0$ and $d<2(2\gamma-1)$;
$s=2$) and for a general case ($d=(s-2)$; $s<2$) for small frequencies.
In each case, the maxima of perturbed specific mass were found to
be exponentially spaced and their distance ratio $\beta$ was found
to be a constant, depending on characteristics of the disc (and on
the perturbations frequency for the first two cases). Inward and outward
flows of gas appeared with negative and positive radial velocities
between minima and maxima of gas perturbed specific mass, leading
the nebular gas eventually to accumulate in the zones of maxima of
perturbed specific mass. We present here analy1ical solutions for
two other general models for small frequencies of the periodic perturbations.
In section 2, the disc model notations and the equations deduced in
Paper 1 are recalled. The general case for $d=(s-2)/2$ is solved
in section 3. We present in section 4 a method to solve a second general
model for $d=s/(\gamma-1)$, called the \textquotedbl polytropic
model\textquotedbl , and a complete solution is found for the particular
values $\gamma=3/2$, $d=-3$ and $s=-1$. We do not know of any previous
similar general analy1ical resolutions. The conclusions are drawn
in section 5 and are similar to the ones of Paper 1. Both papers are
reworked excerpts of \cite{Pletser1990}.

\section{Model notations\label{sec:2 Model notations}}

In Paper 1 \cite{Paper1}, we considered a gaseous disc of mass $M_{d}$
and of specific mass $\rho_{0}=\rho_{c}R^{d}$ with a sound speed
$c_{0}=c_{c}R^{s/2}$, rotating at a circular velocity $v_{0}=v_{0}(R)$
around a central mass $M^{*}$ ($M^{*}>>M_{d}$)· $R=(r/r_{c})$ is
a dimensionless radial distance and the indexes $0$ and $c$ denote
equilibrium characteristics and reference characteristics at the inner
edge of the disc. Allowing for small radial periodic perturbations
of circular frequency $\omega$ to appear, a wave-like equation was
deduced for the spatial term $D$ of the gas perturbed specific mass,
with the prime sign $"\,^{\prime}"=\partial\,/\partial r$,
\begin{equation}
D^{\prime\prime}+\left(2s+1-\frac{d+s}{\gamma}\right)\frac{D^{\prime}}{R}+\left(B^{2}R^{d+2-s}+\omega^{2}A^{2}R^{2-s}+s\left(s-\frac{d+s}{\gamma}\right)\right)\frac{D}{R^{2}}=0\label{eq:1-1}
\end{equation}
where 
\[
A^{2}=\frac{r_{c}^{2}}{c_{c}^{2}}\,\,\,;\,\,B^{2}=\frac{4\pi G\rho_{c}r_{c}^{2}}{c_{c}^{2}}
\]
are constants. The spatial terms of the associated perturbed radial
velocity $U$ and specific mass flux radial momentum $\Phi$ were
found in function of $D$
\begin{align}
U\left(R\right) & =-\kappa\frac{r_{c}}{\rho_{c}}R^{-\left(d+1\right)}\int D\left(R\right)R\,dR\label{eq:2-1}\\
\Phi\left(R\right) & =r_{c}\rho_{c}R^{d+1}U\left(R\right)=-\kappa r_{c}^{2}\int D\left(R\right)R\,dR\label{eq:3-1}
\end{align}
Two boundary conditions were defined by, first, the gas perturbed
specific mass matching at the disc inner edge, for $R=1$, a certain
value independent from disc physical characteristics (see Paper 1)
and, second, decreasing perturbed specific mass for increasing $R$,
vanishing far away from the central mass.

\section{Solutions for d = (s - 2)/2 \label{sec:3 Solutions for d =00003D (s - 2)/2}}

Searching for solutions yielding annular structures to appear in the
disc, we consider a fourth case where the exponents $d$ and $s$
are linked by the relation $d=(s-2)/2$. The equation \eqref{eq:1-1}
reads
\begin{align}
 & D^{\prime\prime}+\left(4d+5-\frac{3d+2}{\gamma}\right)\frac{D^{\prime}}{R}\nonumber \\
 & +\left(\omega^{2}A^{2}R^{-2d}+B^{2}R^{-d}+2\left(d+1\right)\left(2\left(d+1\right)-\left(\frac{3d+2}{\gamma}\right)\right)\right)\frac{D}{R^{2}}=0\label{eq:4-1}
\end{align}
Substituting the variable $R$ for
\[
z=j\left(\frac{2}{d}\omega AR^{-d}\right)
\]
with $j=\sqrt{-1}$, yields a confluent hypergeometric equation
\begin{align}
 & z^{2}\frac{\partial^{2}D}{\partial z^{2}}-\frac{1}{d}\left(3d+4-\left(\frac{3d+2}{\gamma}\right)\right)z\frac{\partial D}{\partial z}\nonumber \\
 & -\left(\frac{z^{2}}{4}+j\frac{B^{2}}{2d\omega A}z-2\left(\frac{d+1}{d^{2}}\right)\left(2\left(d+1\right)-\left(\frac{3d+2}{\gamma}\right)\right)\right)D=0\label{eq:5-1}
\end{align}
With $d\neq-2\,/\left(3+i\gamma\right)$ for all integers $i$, \eqref{eq:5-1}
has a complex solution \cite{Whittaker-Watson1927,Kamke1943,Jahnke-Emde-Losch1966}
\begin{equation}
D_{C}=R^{-\frac{1}{2}\left(3d+4-\left(\frac{3d+2}{\gamma}\right)\right)}\exp\left(\frac{-z}{2}\right)\left(K_{1}z^{\frac{1}{2}\left(1+\left(\frac{3d+2}{d\gamma}\right)\right)}\phi_{1}+K_{2}z^{\frac{1}{2}\left(1-\left(\frac{3d+2}{d\gamma}\right)\right)}\phi_{2}\right)\label{eq:6-1}
\end{equation}
with $K_{1}$ and $K_{2}$ constants and where $\phi_{1}=\phi\left(a_{1},b_{1},z\right)$
and $\phi_{2}=\phi\left(a_{2},b_{2},z\right)$ are Kummer confluent
hypergeometric function of arguments
\begin{align*}
a_{1} & =\frac{1}{2}\left(1+\frac{3d+2}{d\gamma}\right)+jy\,\,\,;\,\,b_{1}=1+\frac{3d+2}{d\gamma}\\
a_{2} & =\frac{1}{2}\left(1-\frac{3d+2}{d\gamma}\right)+jy\,\,\,;\,\,b_{1}=1-\frac{3d+2}{d\gamma}
\end{align*}
where
\[
y=\frac{B^{2}}{d\omega A\left(\frac{3d+2}{\gamma}-\left(3d+4\right)\right)}
\]
For $b_{1}$ and $b_{2}$ non null and different from negative integers,
the Kummer function $\phi$ expands for both sets of arguments as
\cite{Slater1965} 
\begin{equation}
\phi\left(a,b,z\right)=\Gamma\left(\nu\right)\left(\frac{z}{4}\right)^{-\nu}\exp\left(\frac{z}{2}\right)\sum_{k=0}^{\infty}\frac{\left(-1\right)^{k}\left(2\nu\right)_{k}\left(b-2a\right)_{k}}{k!\left(b\right)_{k}}\,I_{\nu+k}\left(\frac{z}{2}\right)\label{eq:7-1}
\end{equation}
where $\Gamma$ is the Legendre Gamma function, $(x)_{k}$ are Pochhammer
polynomials,
\[
(x)_{k}=\prod_{q=0}^{k-1}(x+q)\,\,\,;\,\,(x)_{0}=1
\]
and $I_{\nu+k}$ is the complex valued hyperbolic Bessel function
of order $\left(\nu+k\right)$, with $\nu$ either of
\begin{align*}
\nu_{1} & =\left(b_{1}-a_{1}-\frac{1}{2}\right)=\frac{3d+2}{2d\gamma}-jy\\
\nu_{2} & =\left(b_{2}-a_{2}-\frac{1}{2}\right)=-\frac{3d+2}{2d\gamma}+jy
\end{align*}
Developing the complex coefficients $H_{k}$ of $I_{\nu+k}$ in \eqref{eq:7-1}
as in Appendix A \eqref{sec:Appendix-A}, $D_{C}$ \eqref{eq:6-1}
becomes
\begin{align}
D_{C} & =CR^{-2\left(d+1\right)+\left(\frac{3d+2}{2\gamma}\right)}\frac{z}{4}jy\left(K_{1}C_{1}\sum_{k=0}^{\infty}\left(H_{1k}I_{\nu_{1}+k}\left(\frac{z}{2}\right)\right)\right.\nonumber \\
 & \left.+K_{2}C_{2}\sum_{k=0}^{\infty}\left(H_{2k}I_{\nu_{2}+k}\left(\frac{z}{2}\right)\right)\right)\label{eq:8-1}
\end{align}
where
\[
C=\sqrt{\frac{\omega A}{d}}\exp\left(j\frac{\pi}{4}\right)\,\,\,;\,\,C_{1}=2^{1+\frac{3d+2}{d\gamma}}\Gamma\left(\nu_{1}\right)\,\,\,;\,\,C_{2}=2^{1-\frac{3d+2}{d\gamma}}\Gamma\left(\nu_{2}\right)
\]
are complex constants. Simple analytical expressions of zeros and
extrema of the real part of $D_{C}$ \eqref{eq:8-1} were not found.
However, for small arguments $z$, $\left(\left|z\right|/2\right)<<1$,
i.e. for small frequencies
\begin{equation}
\omega<<\left|d\right|\frac{c_{c}}{r_{c}}R^{-\left|d\right|}\label{eq:9-1}
\end{equation}
where vertical bars denote the absolute value, the terms other than
the first in the convergent series of \eqref{eq:7-1} can be neglected.
By the multiplication theorem \cite{Olver1965}, the hyperbolic Bessel
function reduces then to
\begin{equation}
I_{\nu}\left(\frac{z}{2}\right)\approx\left(\frac{z}{4}\right)^{-\nu}\sum_{m=0}^{\infty}\sum_{n=0}^{\infty}\frac{\left(-1\right)^{m}}{m!n!\Gamma\left(\nu-m+n+1\right)}=\left(\frac{z}{4}\right)^{-\nu}\frac{G\left(\nu\right)}{\nu\Gamma\left(\nu\right)}\label{eq:10-1}
\end{equation}
where terms of second order were neglected in front of unity and where
$G(\nu)$ is a complex function developed in Appendix B \eqref{sec:Appendix-B}.
The relation \eqref{eq:7-1} reads now
\begin{equation}
\phi\left(a,b,z\right)=\frac{G\left(\nu\right)}{\nu}\left(\frac{z}{4}\right)^{-2\nu}\exp\left(\frac{z}{2}\right)\label{eq:11-1}
\end{equation}
The complex solution \eqref{eq:8-1}, written for the variable $R$,
becomes
\begin{equation}
D_{C}=\left(K_{1}\varLambda R^{-2\left(d+1\right)+\frac{3d+2}{\gamma}}+K_{2}MR^{-2\left(d+1\right)}\right)\exp\left(2jy\ln\left(\frac{2}{\left|d\right|}\omega AR^{-d}\right)\right)\label{eq:12-1}
\end{equation}
where $\varLambda=\left|\varLambda\right|\exp\left(j\lambda\right)$
and $M=\left|M\right|\exp\left(j\mu\right)$ are complex constants
depending on $y$, $\nu$ and $G(\nu)$ (see Appendix B). The real
part of \eqref{eq:12-1} reads as the sum of two terms
\begin{align}
D_{R} & =K_{1}\left|\varLambda\right|R^{-2\left(d+1\right)+\frac{3d+2}{\gamma}}\cos\left(2y\ln\left(\frac{2}{\left|d\right|}\omega AR^{-d}\right)+\lambda\right)\nonumber \\
 & +K_{2}\left|M\right|R^{-2\left(d+1\right)}\cos\left(2y\ln\left(\frac{2}{\left|d\right|}\omega AR^{-d}\right)+\mu\right)\label{eq:13-1}
\end{align}
Due to the second boundary condition (decrease of $D$ for increasing
$R$), either the first or the second or both terms of \eqref{eq:13-1}
should be considered for the general solution, depending on the respective
values of $d$ and $\gamma$ as indicated in Table 1.

\begin{table}

\caption{Terms of solution $D$ \eqref{eq:13-1} for respective values of $d$
and $\gamma$}

\centering{}%
\begin{tabular}{|c||c|c|c|c|c|}
\hline 
 & \multicolumn{2}{c|}{$\gamma<\frac{3}{2}$} & $\gamma=\frac{3}{2}$ & \multicolumn{2}{c|}{$\gamma>\frac{3}{2}$}\tabularnewline
\cline{2-6} \cline{3-6} \cline{4-6} \cline{5-6} \cline{6-6} 
 & $d<\frac{2\left(\gamma-1\right)}{3-2\gamma}$ & $d\geq\frac{2\left(\gamma-1\right)}{3-2\gamma}$ & all values of $d$ & $d\leq\frac{2\left(\gamma-1\right)}{3-2\gamma}$ & $d>\frac{2\left(\gamma-1\right)}{3-2\gamma}$\tabularnewline
\hline 
\hline 
$d>-1$ & Both parts & $K_{1}=0$ & Both parts & Not Applicable & Both parts\tabularnewline
\hline 
$d\leq-1$ & $K_{2}=0$ & Not Applicable & $K_{2}=0$ & No decrease & $K_{2}=0$\tabularnewline
\hline 
\end{tabular}
\end{table}

Without loss of generality in the resolution, we consider from now
on only the case $s<0$ and $d<-1$, yielding $K_{2}=0$ in \eqref{eq:13-1},
the other constant $K_{1}$ being fully determined by the first boundary
condition. 

The perturbed radial velocity $U$ and the specific mass flux momentum
$\varPhi$ read, from \eqref{eq:2-1} and \eqref{eq:3-1}, 
\begin{align}
U & =-\kappa K_{1}C_{3}\frac{r_{c}}{\rho_{c}}R^{-\left(3d+1\right)+\frac{3d+2}{\gamma}}\sin\left(2y\ln\left(\frac{2}{\left|d\right|}\omega AR^{-d}\right)+\lambda+\zeta\right)\label{eq:14-1}\\
\varPhi & =-\kappa K_{1}C_{3}r_{c}^{2}R^{-2d+\frac{3d+2}{\gamma}}\sin\left(2y\ln\left(\frac{2}{\left|d\right|}\omega AR^{-d}\right)+\lambda+\zeta\right)\label{eq:15}
\end{align}
with
\[
C_{3}=\frac{\left|\varLambda\right|}{\sqrt{\left(\frac{3d+2}{\gamma}-2d\right)^{2}+\left(-2dy\right)^{2}}}\,\,\,;\,\,\zeta=\arctan\left(\frac{\frac{3d+2}{\gamma}-2d}{-2dy}\right)
\]
The extrema (minima and maxima) of $D_{R}$ are found from
\begin{equation}
D_{R}^{\prime}=-K_{1}C_{4}R^{-2\left(d+3\right)+\frac{3d+2}{\gamma}}\sin\left(2y\ln\left(\frac{2}{\left|d\right|}\omega AR^{-d}\right)+\lambda+\xi\right)=0\label{eq:16-1}
\end{equation}
with
\[
C_{4}=\left|\varLambda\right|\sqrt{\left(\frac{3d+2}{\gamma}-2\left(d+1\right)\right)^{2}+\left(2dy\right)^{2}}\,\,\,;\,\,\xi=\arctan\left(\frac{\frac{3d+2}{\gamma}-2\left(d+1\right)}{2dy}\right)
\]
The zeros of $D_{R}$ \eqref{eq:13-1}, $U$ \eqref{eq:14-1}, $\varPhi$
\eqref{eq:15} and $D_{R}^{\prime}$ \eqref{eq:16-1} read in this
fourth case
\begin{equation}
R=\alpha_{4}\left(\beta_{4}^{\backprime}\right)^{n}\,\,\,;\,\,\alpha_{4}=\left(\frac{2}{\left|d\right|}\omega A\right)^{\frac{1}{d}}\exp\left(\frac{\lambda+\varphi_{4}}{2\left|d\right|y}\right)\,\,\,;\,\,\beta_{4}^{\backprime}=\exp\left(\frac{\pi}{2\left|d\right|y}\right)\label{eq:17-1}
\end{equation}
$n$ being non-negative integers and $\varphi_{4}=\left(\pi/2\right)$
for$D_{R}$, $\varphi_{4}=\zeta$ for $U$ and $\varPhi$, and $\varphi_{4}=\xi$
for $D_{R}^{\prime}$. Provided that $\omega A$ is small enough,
within the condition \eqref{eq:9-1}, one has $y>>1$, yielding $\zeta<<1$
and $\xi<<1$. The initial phase between $D$ and $U$ is $\left(\pi/2\right)-\zeta\approx\left(\pi/2\right)$,
while the initial phase between $U$ (or $\varPhi$) and $D^{\prime}$
is $(\xi-\zeta)\approx0$. 

The distances ratio of two successive maxima of $D$ is
\begin{equation}
\beta_{4}=\left(\beta_{4}^{\backprime}\right)^{2}=\exp\left(\frac{\left(3d+4-\frac{3d+2}{\gamma}\right)\omega c_{c}}{4\pi G\rho_{c}r_{c}}\right)\label{eq:18-1}
\end{equation}
which is a real constant depending on the perturbations circular frequency
$\omega$ and the disc reference characteristics. The period of the
perturbations must be larger than a minimum value
\begin{equation}
P_{m}=\frac{2\pi}{\left|d\right|}\frac{r_{c}}{c_{c}}R_{max}^{d}\label{eq:19-1}
\end{equation}
deduced from the condition \eqref{eq:9-1} applied to the whole range
of radial distances of the disc ($R_{max}$ is the ratio of the outer
and inner radii of the disc).

\section{Solution for the polytropic case\label{sec:4 Solution for the polytropic case}}

\subsection{General formulation\label{subsec:4.1 General-formulation}}

In the previous section and in Paper 1, we considered the exponents
$d$ and $s$ taking particular values or linked by non-causal relations.
However, a relation between $d$ and $s$ can be found if one considers
that the specific mass and sound speed are fully governed by polytropic
processes in the disc. Considering the two polytropic relations between
the pressure $p$, the specific mass $\rho$ and the sound speed $c$
\begin{equation}
p=\frac{c^{2}\rho}{\gamma}\,\,\,;\,\,p\rho^{-\gamma}=\text{constant}\label{eq:20-1}
\end{equation}
one has successively, with the power law radial distributions $\rho_{0}=\rho_{c}R^{d}$
and $c_{0}^{2}=c_{c}^{2}R^{s}$,
\begin{align}
 & c^{2}\rho^{1-\gamma}=c_{c}^{2}\rho_{c}^{1-\gamma}R^{s+d\left(1-\gamma\right)}=\text{constant}\label{eq:21-1}\\
 & s+d\left(1-\gamma\right)=0\,\,\,\text{or}\,\,\gamma=1+\frac{s}{d}\label{eq:22-1}
\end{align}
Replacing $\gamma$ for $s$ and $d$ in the equation \eqref{eq:1-1}
yields
\begin{equation}
D^{\prime\prime}+\left(2s+1-d\right)\frac{D^{\prime}}{R}+\left(B^{2}R^{d+2-s}+\omega^{2}A^{2}R^{2-s}+s\left(s-d\right)\right)\frac{D}{R^{2}}=0\label{eq:23-1}
\end{equation}
which becomes a differential Schrödinger type equation by posing
\[
D=\frac{Y}{\sqrt{R^{2s+1-d}}}
\]
yielding
\begin{equation}
Y^{\prime\prime}+\left(B^{2}R^{d-s}+\omega^{2}A^{2}R^{-s}-\left(\frac{d^{2}-1}{4}\right)R^{-2}\right)Y=0\label{eq:24-1}
\end{equation}
An approximate solution to this equation can be found by the Wentzel-Kramers-Brillouin
(WKB) theory \cite{Bender-Orszag1978}. Considering the case of small
frequencies such as
\begin{equation}
\omega A<<B^{2}\,\,\,\text{or}\,\,\omega<<4\pi G\frac{\rho_{c}r_{c}}{c_{c}}\label{eq:25-1}
\end{equation}
one poses $\epsilon=\omega AB^{-2}$ with $\epsilon<<1$. The equation
\eqref{eq:24-1} reads then
\begin{equation}
\epsilon^{2}Y^{\prime\prime}=\left(-\omega^{2}A^{2}\epsilon^{2}R^{-s}-\omega A\epsilon R^{d-s}+\epsilon^{2}\left(\frac{d^{2}-1}{4}\right)R^{-2}\right)Y=Q\left(R\right)Y\label{eq:26-1}
\end{equation}
Let us consider the three following functions of $R$
\begin{equation}
S_{0}\left(R\right)=\intop^{R}\sqrt{Q(x)}dx;\,S_{1}\left(R\right)=-\frac{\ln\left(Q\left(R\right)\right)}{4};\,S_{2}\left(R\right)=\intop^{R}\left(\frac{QQ^{\prime\prime}-\frac{5}{4}\left(Q^{\prime}\right)^{2}}{8Q^{\frac{5}{2}}}\right)dx\label{eq:27-1}
\end{equation}
where the first two functions are referred to respectively as the
eikonal function and the transport function and where $Q^{\prime}=dQ(x)/dx$. 

If $Q(R)\neq0$ in the range of interest of $R$ (i.e., $1\leq R\leq R_{max}$)
and under the conditions
\begin{equation}
S_{1}\left(R\right)<<\frac{S_{0}\left(R\right)}{\epsilon}\,\,\,;\,\,\epsilon S_{2}\left(R\right)<<S_{1}\left(R\right)\,\,\,;\,\,\epsilon S_{2}\left(R\right)<<1\label{eq:28-1}
\end{equation}
the leading orders in the WKB physical optics approximation to the
exact solutions in $Y$ and $D$ reads generally
\begin{align}
Y\left(R\right) & =K_{3}\exp\left(\frac{S_{0}\left(1,R\right)}{\epsilon}+S_{1}\left(R\right)\right)+K_{4}\exp\left(-\frac{S_{0}\left(1,R\right)}{\epsilon}+S_{1}\left(R\right)\right)\label{eq:29-1}\\
D\left(R\right) & =R^{-\left(2s+1-d\right)/2}\left(K_{3}\exp\left(\frac{S_{0}\left(1,R\right)}{\epsilon}+S_{1}\left(R\right)\right)\right.\nonumber \\
 & \left.+K_{4}\exp\left(-\frac{S_{0}\left(1,R\right)}{\epsilon}+S_{1}\left(R\right)\right)\right)\label{eq:30-1}
\end{align}
with $K_{3}$ and $K_{4}$ constants determined by the boundary conditions
and where $S_{0}\left(1,R\right)$ is the eikonal function on the
interval $\left[1,R\right]$. Strictly speaking, the above equality
sign should be replaced by an asymptotic equality sign. The eikonal
function $S_{0}\left(1,R\right)$ reads
\begin{align}
S_{0}\left(1,R\right)=\intop_{1}^{R}\sqrt{Q\left(x\right)}dx & =j\omega A\epsilon\intop_{1}^{R}\sqrt{x^{-s}+\frac{1}{\omega A\epsilon}x^{d-s}-\left(\frac{d^{2}-1}{4\omega^{2}A^{2}}\right)x^{-2}}dx\nonumber \\
 & =j\omega A\epsilon\intop_{1}^{R}\frac{\sqrt{T\left(x\right)}}{x}dx\label{eq:31-1}
\end{align}
with
\begin{equation}
T\left(x\right)=x^{2-s}+\frac{1}{\omega A\epsilon}x^{2+d-s}-\left(\frac{d^{2}-1}{4\omega^{2}A^{2}}\right)\label{eq:32-1}
\end{equation}
The integral \eqref{eq:31-1} has to be evaluated for specific values
of $d$ and $s$. This evaluation involves most of the time elliptic
integrals, which makes it uneasy. 

\subsection{Solution for $\gamma=3/2$, d = -2 and s = -1\label{subsec:4.2 4.2 Solution for =00005Cgamma=00003D3/2, d =00003D -2 and s =00003D -1}}

In most nebula models, the gas specific mass and sound speed are decreasing
outward from the central body, with the exponents $d$ and $s$ taking
negative values and $s$ usually in the order of or close to $-1$.
We consider here the particular polytropic case with $s=-1$ and $d=-2$,
yielding $\gamma=3/2$. Cases for other values of $d$, $s$ and $\gamma$
can be solved similarly. The integral \eqref{eq:31-1} reads
\begin{equation}
S_{0}\left(1,R\right)=j\left(\frac{\omega A\epsilon}{3}I_{1}+\frac{2}{3}I_{2}-\frac{3\epsilon}{4\omega A}I_{3}\right)\label{eq:33-1}
\end{equation}
with
\begin{equation}
I_{1}=\intop_{1}^{R}\frac{3x^{2}+\frac{1}{\omega A\epsilon}}{\sqrt{T\left(x\right)}}dx\,\,\,;\,\,I_{2}=\intop_{1}^{R}\frac{dx}{\sqrt{T\left(x\right)}}\,\,\,;\,\,I_{3}=\intop_{1}^{R}\frac{dx}{x\sqrt{T\left(x\right)}}\label{eq:33-2}
\end{equation}
The cubic trinomial $T\left(x\right)$ \eqref{eq:32-1} has a single
real root and, neglecting terms in $\epsilon^{2}$ and of higher order,
it becomes
\begin{equation}
T\left(x\right)=\left(x-\frac{3\epsilon}{4\omega A}\right)\left(x^{2}+\frac{3\epsilon}{4\omega A}x+\frac{1}{\omega A\epsilon}\right)\label{eq:34-1}
\end{equation}
Posing
\[
C_{5}=\sqrt{\frac{1}{\omega A\epsilon}+\frac{9}{8}\left(\frac{\epsilon}{\omega A}\right)^{2}}
\]
one substitutes $\left(x-\frac{3\epsilon}{4\omega A}\right)$ for
$C_{5}z$ in integral $I_{2}$ and for $C_{5}z^{2}$ in integral $I_{3}$
in \eqref{eq:33-2}. The integrals in \eqref{eq:33-2} are evaluated
under the two following conditions
\begin{equation}
\frac{\omega A}{B}R<<1\,\,\,;\,\,B^{2}R>>\frac{3}{4}\label{eq:35-1}
\end{equation}
in the range $1\leq R\leq R_{max}$, showing also that the real root
of the trinomial $T\left(x\right)$ \eqref{eq:34-1} is outside the
range of interest of $R$, fulfilling the condition $Q(R)\neq0$.
The other conditions \eqref{eq:28-1} of application of the WKB theory
are verified in Appendix C. 

Under the above two conditions, $I_{2}$ and $I_{3}$ become \cite{Gradshteyn-Ryzhik1965}
\begin{align}
I_{2} & =\frac{1}{\sqrt{C_{2}}}\left(F\left(\boldsymbol{\varphi}\left(R\right),k\right)-F\left(\boldsymbol{\varphi}\left(1\right),k\right)\right)\label{eq:36-2}\\
I_{3} & =\frac{1}{\sqrt{\left(C_{2}\right)^{3}}}\left(\left(F\left(\boldsymbol{\varphi}\left(R\right),k\right)-2E\left(\boldsymbol{\varphi}\left(R\right),k\right)\right)-\left(F\left(\boldsymbol{\varphi}\left(1\right),k\right)-2E\left(\boldsymbol{\varphi}\left(1\right),k\right)\right)\right)\nonumber \\
 & -\frac{2}{C_{2}}\left(\frac{1}{R-\frac{3\epsilon}{4\omega A}+C_{2}}\sqrt{\frac{R^{2}+\frac{3\epsilon}{4\omega A}R+\frac{1}{\omega A\epsilon}}{R-\frac{3\epsilon}{4\omega A}}}\right.\nonumber \\
 & \left.-\left(\frac{1}{1-\frac{3\epsilon}{4\omega A}+C_{2}}\sqrt{\frac{1+\frac{3\epsilon}{4\omega A}+\frac{1}{\omega A\epsilon}}{1-\frac{3\epsilon}{4\omega A}}}\right)\right)\label{eq:37-1}
\end{align}
where $F$ and $E$ are the incomplete elliptic integrals of the first
and second kinds of argument and modulus
\begin{align}
 & \boldsymbol{\varphi}\left(R\right)=2\arctan\left(\frac{1}{C_{2}}\sqrt{R-\frac{3\epsilon}{4\omega A}}\right)\approx2\arctan\left(\sqrt{\frac{\omega A}{B}R}\right)\label{eq:38-1}\\
 & k=\sqrt{\frac{1}{2}-\frac{9\epsilon}{16\omega A\sqrt{C_{2}}}}\approx\frac{1}{\sqrt{2}}\label{eq:39-1}
\end{align}
where the conditions \eqref{eq:35-1} were used, yielding also $C_{2}=\frac{B}{\omega A}$.
Replacing in \eqref{eq:37-1}, \eqref{eq:38-1} and in \eqref{eq:33-2},
\eqref{eq:33-1} yields eventually
\begin{equation}
S_{0}\left(1,R\right)=j\epsilon\left(s_{0R}-s_{01}\right)\label{eq:40-1}
\end{equation}
with
\begin{align}
s_{0R} & =\left(\frac{2}{3}\sqrt{\frac{B^{3}}{\omega A}}-\frac{3}{4}\sqrt{\frac{\omega A}{B^{3}}}\right)F\left(\boldsymbol{\varphi}\left(R\right),k\right)+\frac{3}{2}\sqrt{\frac{\omega A}{B^{3}}}E\left(\boldsymbol{\varphi}\left(R\right),k\right)\nonumber \\
 & +\frac{\sqrt{\left(\omega A\right)^{2}R^{3}+B^{2}R-\frac{3}{4}}\left(\frac{2}{3}+\frac{3}{2}\frac{B^{3}}{\omega A}\right)}{\left(B^{2}R-\frac{3}{4}\right)\left(B^{2}R-\frac{3}{4}+\frac{B^{3}}{\omega A}\right)}\label{eq:41-1}
\end{align}
and a similar relation for $s_{01}$ with $R=1$ . 

The incomplete elliptic integrals $F$ and $E$ are evaluated after
an ascending Landen transformation, yielding the new argument and
modulus and the transformed expressions of $F$ and $E$ to be
\begin{align}
\boldsymbol{\varphi}_{t} & =\frac{1}{2}\left(\boldsymbol{\varphi}+\arcsin\left(k\sin\boldsymbol{\varphi}\right)\right)\approx\frac{1}{2}\arcsin\left(2\left(1+\frac{1}{\sqrt{2}}\right)\sqrt{\frac{\omega A}{B}R}\right)\label{eq:42-1}\\
k_{t} & =\frac{2\sqrt{k}}{1+k}\approx\frac{2^{\frac{5}{4}}}{1+\sqrt{2}}\approx0.9852\label{eq:43-1}\\
F\left(\boldsymbol{\varphi}\left(R\right),k\right) & =\frac{2}{1+k}F\left(\boldsymbol{\varphi_{t}}\left(R\right),k_{t}\right)\label{eq:44-1}\\
E\left(\boldsymbol{\varphi}\left(R\right),k\right) & =\left(1+k\right)E\left(\boldsymbol{\varphi_{t}}\left(R\right),k_{t}\right)+\left(1-k\right)F\left(\boldsymbol{\varphi_{t}}\left(R\right),k_{t}\right)\nonumber \\
 & -\frac{\left(1+k\right)\tan\boldsymbol{\varphi}}{2+\frac{\sec\boldsymbol{\varphi}\left(1+\frac{1}{k^{2}}\right)\sqrt{1-k^{2}}}{\cos\boldsymbol{\varphi}+\sqrt{\frac{1}{k^{2}}-\sin^{2}\boldsymbol{\varphi}}}}\label{eq:45-1}
\end{align}
As $k_{t}$ is close to unity, one can use the expansions \cite{Gradshteyn-Ryzhik1965}
\begin{align}
F\left(\boldsymbol{\varphi_{t}}\left(R\right),k_{t}\right) & =\frac{2}{\pi}\boldsymbol{K^{\prime}}\ln\left(\tan\left(\frac{\boldsymbol{\varphi}_{t}}{2}+\frac{\pi}{4}\right)\right)-\sin\boldsymbol{\varphi}_{t}\sec^{2}\boldsymbol{\varphi}_{t}\left(a_{0}-\frac{2}{3}a_{1}\tan^{2}\boldsymbol{\varphi}_{t}+...\right)\label{eq:46-1}\\
E\left(\boldsymbol{\varphi_{t}}\left(R\right),k_{t}\right) & =\frac{2}{\pi}\boldsymbol{E^{\prime}}\ln\left(\tan\left(\frac{\boldsymbol{\varphi}_{t}}{2}+\frac{\pi}{4}\right)\right)+\sin\boldsymbol{\varphi}_{t}\sec^{2}\boldsymbol{\varphi}_{t}\left(b_{0}-\frac{2}{3}b_{1}\tan^{2}\boldsymbol{\varphi}_{t}+...\right)\label{eq:47-1}
\end{align}
where
\[
\boldsymbol{K^{\prime}}=\boldsymbol{K}\sqrt{1-k_{t}^{2}}\,\,\,;\,\,\boldsymbol{E^{\prime}}=\boldsymbol{E}\sqrt{1-k_{t}^{2}}
\]
are the complete integrals of the first and second kinds and $a_{0},a_{1},...,b_{0},b_{1},...$
are decreasing coefficients, functions of $k_{t}$. As $k_{t}\approx0.9852$,
$\boldsymbol{K^{\prime}}\approx\boldsymbol{E^{\prime}}\approx\pi/2$
within $7.5\times10^{-3}$, $a_{0}\approx7.5\times10^{-3}$, $a_{1}\approx10^{-4}$,
$b_{0}\approx-7.5\times10^{-3}$, etc ... , yielding
\begin{equation}
F\left(\boldsymbol{\varphi_{t}}\left(R\right),k_{t}\right)\approx E\left(\boldsymbol{\varphi_{t}}\left(R\right),k_{t}\right)\approx\ln\left(\frac{1}{2}\left(1+\frac{1}{\sqrt{2}}\right)\sqrt{\frac{\omega A}{B}R}\right)\label{eq:48-1}
\end{equation}
Replacing in \eqref{eq:44-1}, \eqref{eq:45-1} and \eqref{eq:41-1}
and using conditions \eqref{eq:35-1} to neglect small terms, it yields
\begin{equation}
s_{0R}=W\ln\left(\frac{1}{2}\left(1+\frac{1}{\sqrt{2}}\right)\sqrt{\frac{\omega A}{B}R}\right)+\frac{2}{3}\sqrt{\left(\omega A\right)^{2}R^{3}+B^{2}R-\frac{3}{4}}\label{eq:49-1}
\end{equation}
and similarly for $s_{01}$ for $R=1$, with $W=W(\omega,A,B)$
\begin{equation}
W=\frac{8}{3}\left(1-\frac{1}{\sqrt{2}}\right)\sqrt{\frac{B^{3}}{\omega A}}+3\sqrt{\frac{\omega A}{2B^{3}}}\approx\frac{8}{3}\left(1-\frac{1}{\sqrt{2}}\right)\sqrt{\frac{B^{3}}{\omega A}}\label{eq:50-1}
\end{equation}
under the conditions \eqref{eq:35-1}. 

The complex perturbed specific mass $D_{C}$ \eqref{eq:30-1} and
its real part read
\begin{align}
D_{C} & =\frac{B}{\sqrt{j\omega A}\sqrt[4]{\left(\omega A\right)^{2}R^{3}+B^{2}R-\frac{3}{4}}}\left(K_{3}\exp\left(j\left(s_{0R}-s_{01}\right)\right)\right.\nonumber \\
 & \left.+K_{4}\exp\left(-j\left(s_{0R}-s_{01}\right)\right)\right)\label{eq:51-1}\\
D_{R} & =K_{5}\frac{B}{\sqrt{\omega A}\sqrt[4]{\left(\omega A\right)^{2}R^{3}+B^{2}R-\frac{3}{4}}}\cos\left(s_{0R}-s_{01}-\kappa\right)\label{eq:52-1}
\end{align}
with
\[
K_{5}=\sqrt{K_{3}^{2}+K_{4}^{2}}\,\,\,;\,\,\kappa=\arctan\left(\frac{K_{3}-K_{4}}{K_{3}+K_{4}}\right)
\]
The extrema of $D_{R}$ are solutions of
\begin{equation}
D_{R}^{\prime}=-K_{5}C_{6}\frac{1}{\sqrt[4]{\left(\omega A\right)^{2}R^{3}+B^{2}R-\frac{3}{4}}}\sin\left(s_{0R}-s_{01}-\kappa+\tau\right)=0\label{eq:53-1}
\end{equation}
with
\[
C_{6}=\frac{4}{3}\left(1-\frac{1}{\sqrt{2}}\right)\frac{B^{\frac{5}{2}}}{\omega A}\,\,\,;\,\,\tau=\arctan\left(\frac{3}{8}\left(1+\frac{1}{\sqrt{2}}\right)\sqrt{\frac{\omega A}{B^{3}}}\right)
\]
where the conditions \eqref{eq:35-1} were used to neglect small terms.
The zeros and extrema of $D_{R}$ are given by
\begin{align}
 & \frac{1}{2}\left(1+\frac{1}{\sqrt{2}}\right)\sqrt{\frac{\omega A}{B}R}\exp\left(\frac{2}{3}\frac{\sqrt{\left(\omega A\right)^{2}R^{3}+B^{2}R-\frac{3}{4}}}{W}\right)\nonumber \\
 & =\exp\left(\frac{s_{01}+\varphi_{5}+\kappa}{W}\right)\exp\left(\frac{\pi n}{W}\right)\label{eq:54-1}
\end{align}
where $n$ are non-negative integers and $\varphi_{5}=\pi/2$ for
$D_{R}$ and $\varphi_{5}=-\tau$ for $D_{R}^{\prime}$. The exponential
term in the above left hand side part reduces to
\begin{equation}
\exp\left(\frac{2}{3}\frac{\sqrt{\left(\omega A\right)^{2}R^{3}+B^{2}R-\frac{3}{4}}}{W}\right)\approx\exp\left(\frac{1}{2}\left(1+\frac{1}{\sqrt{2}}\right)\sqrt{\frac{\omega A}{B}R}\right)\label{eq:55-1}
\end{equation}
under the conditions \eqref{eq:35-1}, and it can be neglected when
multiplied by its argument, small by \eqref{eq:35-1}. The zeros and
extrema of $D_{R}$ are then given in good approximation by
\begin{equation}
R=\alpha_{5}\left(\beta_{5}^{\backprime}\right)^{n};\,\alpha_{5}=16\left(\frac{3}{2}-\sqrt{2}\right)\frac{B}{\omega A}\exp\left(\frac{2\left(s_{01}+\varphi_{5}+\kappa\right)}{W}\right);\,\beta_{5}^{\backprime}=\exp\left(\frac{2\pi}{W}\right)\label{eq:56-1}
\end{equation}
The perturbed radial velocity $U$ and the specific mass flux radial
momentum $\varPhi$ are found from \eqref{eq:2-1} and \eqref{eq:3-1},
with \eqref{eq:52-1}. However, their evaluation requires the resolution
of a new elliptic integral. To avoid this and as there are no zeros
due to the transport function $S_{1}(R)$ (in the 4-th root of the
trinomial term in $R$) in \eqref{eq:52-1}, we evaluate $U$ and
$\varPhi$ by neglecting $S_{1}(R)$. This is the geometrical optics
approximation, which gives the most rapidly varying component (controlling
factor) of the leading behaviour of the exact solution. In the geometrical
optics (g.o.) approximation, the real part of the complex perturbed
specific mass \eqref{eq:52-1} reduces then to
\begin{equation}
D_{R\,g.o.}=K_{5}\frac{B}{\sqrt{\omega A}}\cos\left(s_{0R}-s_{01}-\kappa\right)\label{eq:57-1}
\end{equation}
Under the conditions \eqref{eq:35-1}, the term $s_{0R}$ \eqref{eq:49-1}
can be written approximately
\begin{equation}
s_{0R}\approx W\left(\ln\left(\frac{1}{2}\left(1+\frac{1}{\sqrt{2}}\right)\sqrt{\frac{\omega A}{B}R}\right)+\frac{1}{2}\left(1+\frac{1}{\sqrt{2}}\right)\sqrt{\frac{\omega A}{B}R}\right)\label{eq:58-1}
\end{equation}
as was already done in \eqref{eq:55-1}. The radial velocity $U$
and the specific mass flux radial momentum $\varPhi$ read then in
the geometrical optics approximation
\begin{align}
U_{g.o.} & =-\kappa K_{5}C_{7}\frac{r_{c}}{\rho_{c}}R^{3}\sin\left(s_{0R}-s_{01}-\kappa+\sigma\right)\label{eq:59-1}\\
\varPhi_{g.o.} & =-\kappa K_{5}C_{7}r_{c}^{2}R^{2}\sin\left(s_{0R}-s_{01}-\kappa+\sigma\right)\label{eq:60-1}
\end{align}
with
\[
C_{7}=\frac{2B}{\sqrt{\omega A\left(W^{2}+16\right)}}\,\,\,;\,\,\sigma=\arctan\left(3\left(1+\frac{1}{\sqrt{2}}\right)\sqrt{\frac{\omega A}{B^{3}}}\right)
\]
showing that their zeros are given like the zeros and extrema of $D_{R}$
with $\beta_{5}^{\backprime}$ \eqref{eq:56-1} and $\varphi_{5}=-\sigma$
in $\alpha_{5}$ \eqref{eq:56-1}. 

By the condition \eqref{eq:25-1}, one has $\sigma<<1$ and $\tau<<1$.
The initial phase between $D$ and $U$ (or $\varPhi$) is $\left(\pi/2\right)-\sigma\approx\pi/2$,
while the initial phase between $U$ (or $\varPhi$) and $D^{\prime}$
is $\left(\tau-\sigma\right)\approx0$. 

Let us note that, in the physical optics approximation, retaining
the term $S_{1}(R)$ in $D_{R}$ would change the amplitudes of $U$
and $\varPhi$ (by the addition of decreasing terms in $R$ in front
of the sin function) and it would change the coefficient of $\sqrt{\omega A/B^{3}}$
and add negligible terms in the argument of the $\arctan$ of $\sigma$
. But the distance ratio of two successive zeros of $U$ (or $\varPhi$)
is unaffected and is still given by $\beta_{5}^{\backprime}$ \eqref{eq:56-1}
in both WKB approximations. 

The distances ratio of two successive maxima of $D$ is given by
\begin{equation}
\beta_{5}=\left(\beta_{5}^{\backprime}\right)^{2}\label{eq:62-1}
\end{equation}
with
\begin{align}
\beta_{5} & =\exp\left(\frac{\pi}{\frac{2}{3}\left(1-\frac{1}{\sqrt{2}}\right)\frac{1}{\sqrt{\omega}}\left(4\pi G\rho_{c}\right)^{\frac{3}{4}}\frac{r_{c}}{c_{c}}+\frac{3}{4}\sqrt{\frac{\omega}{2}}\left(4\pi G\rho_{c}\right)^{-\frac{3}{4}}\frac{c_{c}}{cr_{c}}}\right)\nonumber \\
 & \approx\exp\left(\frac{3\pi\left(1+\frac{1}{\sqrt{2}}\right)c_{c}\sqrt{\omega}}{r_{c}\left(4\pi G\rho_{c}\right)^{\frac{3}{4}}}\right)\label{eq:63-1}
\end{align}
where the approximated value of W \eqref{eq:50-1} is used in \eqref{eq:63-1}.
The distances ratio $\beta_{5}$ is a constant, function of the perturbations
circular frequency $\omega$ and of the disc reference characteristics.
The period of the small perturbations must be larger than a minimum
value $P_{m}$, which is the greatest of the two values that can be
deduced from the two conditions \eqref{eq:25-1} and \eqref{eq:35-1}
on $\omega$, applied to the whole range of radial distances up to
$R_{max}$, yielding

\begin{equation}
P_{m}=\frac{c_{c}}{2G\rho_{c}r_{c}}\,\,\,\text{or}\,\,\,P_{m}=\sqrt{\frac{\pi}{G\rho_{c}}}R_{max}\label{eq:64-1}
\end{equation}

\section{Conclusions\label{sec:5 Conclusions}}

We have extended the resolution of the wave-like equation of perturbed
specific mass deduced in Paper 1 to two other general cases. The solution
for the \textquotedbl polytropic case\textquotedbl{} could not be
solved generally as one must choose particular values of $d$ and
$s$, fixing the value of the polytropic index $\gamma$. However,
a solution was found in the WKB physical-optics approximation for
an important particular case ($\gamma=3/2$ with $d=-2$ and $s=-1$). 

For the two above cases, conclusions similar to the ones of Paper
1 are reached concerning the functions $D$, $D^{\prime}$, $U$ and
$\varPhi$, namely that, first, $D$ has a sign opposite to the signs
of $D^{\prime}$, $U$ and $\varPhi$; second, the functions $D^{\prime}$,
$U$ and $\varPhi$ are in phase and have an initial phase difference
of approximately $\pi/2$ with respect to the function $D$; third,
the zeros of $U$ corresponds to the extrema of $D$ and vice-versa;
and finally, for increasing $R$, the functions $U$ and $\varPhi$
are positive (respectively negative) between successive minima and
maxima (respectively successive maxima and minima) of $D$. This situation
yields radial outward flows of gas between successive minima and maxima
of $D$ and radial inward flows of gas between successive maxima and
minima of $D$, that would eventually form annular structures of gas,
with axial radii corresponding to the distances of maxima of the gas
perturbed specific mass. Furthermore, the maxima of the gas perturbed
specific mass are found to be exponentially spaced for the two cases
and their distances ratios are constants depending on discs characteristics
and on the circular frequency of the perturbations. These results
can be applied to protoplanetary and proto-satellite discs. 
\begin{acknowledgements}
We thank Dr D. Poelaert for some mathematical advice for Section 3
and Dr T. Dewandre for suggesting the WKB resolution of Section 4
and for valuable discussions.
\end{acknowledgements}

\section*{Appendix A\label{sec:Appendix-A}}

The $k$-th complex coefficient of the hyperbolic Bessel functions
in the series of \eqref{eq:7-1} reads
\begin{equation}
H_{k}=\frac{\left(-1\right)^{k}\left(2\nu\right)_{k}\left(b-2a\right)_{k}}{k!\,\left(b\right)_{k}}\label{eq:A1}
\end{equation}
where $\left(x\right)_{k}$ are Pochhammer polynomials. This expression
can be written $H_{k}=H_{Rk}+jH_{Ik}$ by posing $a=a_{R}+ja_{I}$
and by developing the Pochhammer polynomials, yielding
\begin{equation}
H_{Rk}=T_{k}\left(P_{Rk}Q_{Rk}-P_{Ik}Q_{Ik}\right);\,H_{Ik}=T_{k}\left(P_{Rk}Q_{Ik}-P_{Ik}Q_{Rk}\right)\label{eq:A2}
\end{equation}
with
\begin{align}
P_{Rk} & =\sum_{m=0}^{L_{R}}\left(\sum_{q=2m}^{k}\left(-1\right)^{m+q}\left(_{2m}^{q}\right)S_{k}\left(q\right)a_{R}^{q-2m}a_{I}^{2m}\right)\label{eq:A3}\\
Q_{Rk} & =\sum_{m=0}^{L_{R}}\left(-1\right)^{m}\left|S_{k}\left(2m\right)\right|a_{I}^{2m}\label{eq:A4}\\
P_{Ik} & =\sum_{m=0}^{L_{I}}\left(\sum_{q=2m+1}^{k}\left(-1\right)^{m+q}\left(_{2m+1}^{q}\right)S_{k}\left(q\right)a_{R}^{q-\left(2m+1\right)}a_{I}^{2m+1}\right)\label{eq:A5}\\
Q_{Ik} & =\sum_{m=0}^{L_{I}}\left(-1\right)^{m}\left|S_{k}\left(2m+1\right)\right|a_{I}^{2m+1}\label{eq:A6}\\
T_{k} & =\frac{\left(-1\right)^{k}}{k!\sum_{m=0}^{k}\left|S_{k}\left(m\right)\right|b^{m}}\label{eq:A7}
\end{align}
where $\left(_{2m}^{q}\right)$ are the binomial coefficients, $\left|S_{k}\left(m\right)\right|$
is the absolute value of the Stirling numbers of the first kind and
$L_{R}=L_{I}=k/2$ for $k$ even and $L_{R}=\left(k-1\right)/2$,
$L_{I}=\left(k+1\right)/2$ for $k$ odd.

\section*{Appendix B\label{sec:Appendix-B}}

Writing $\nu=\nu_{R}+j\nu_{I}$, with
\begin{equation}
\nu_{R}=\frac{3d+2}{2d\gamma}\,\,\,;\,\,\nu_{I}=-y=\frac{-B^{2}}{\omega Aa\left(\frac{3d+2}{\gamma}\right)-\left(3d+4\right)}\label{eq:B1}
\end{equation}
the term $G\left(\nu\right)$ in \eqref{eq:10-1} can be written $G\left(\nu\right)=G_{R}+jG_{I}$
with, for $\nu_{1}$,
\begin{align}
G_{1R} & =\sum_{m=0}^{\infty}\frac{1}{m!}\left(\sum_{n=0}^{m}P_{n}+\sum_{n=m+1}^{\infty}\frac{T_{n}}{C_{n}}\right);G_{1I}=\sum_{m=0}^{\infty}\frac{1}{m!}\left(\sum_{n=0}^{m}Q_{n}+\sum_{n=m+1}^{\infty}\frac{W_{n}}{C_{n}}\right)\label{eq:B2}\\
P_{n} & =\frac{1}{n!}\sum_{p=0}^{L_{1}}\sum_{q=2p}^{m-n}\left(-1\right)^{p-n}\left(_{2p}^{q}\right)\left(S_{m-n}\left(q\right)\right)\nu_{R}^{q-2p}\nu_{I}^{2p}\label{eq:B3}\\
Q_{n} & =\frac{1}{n!}\sum_{p=0}^{L_{2}}\sum_{q=2p+1}^{m-n}\left(-1\right)^{p-n}\left(_{2p+1}^{q}\right)\left(S_{m-n}\left(q\right)\right)\nu_{R}^{q-2p-1}\nu_{I}^{2p+1}\label{eq:B4}\\
T_{n} & =\sum_{p=0}^{L_{3}}\sum_{q=2p}^{n-m}\left(-1\right)^{m+p+q}\left(_{2p}^{q}\right)\left(S_{n-m}\left(q\right)\right)\left(\nu_{R}+1\right)^{q-2p}\nu_{I}^{2p}\label{eq:B5}\\
W_{n} & =\sum_{p=0}^{L_{4}}\sum_{q=2p+1}^{n-m}\left(-1\right)^{m+p+q+1}\left(_{2p+1}^{q}\right)\left(S_{n-m}\left(q\right)\right)\left(\nu_{R}+1\right)^{q-2p-1}\nu_{I}^{2p+1}\label{eq:B6}\\
C_{n} & =n!\left(T_{n}^{2}+W_{n}^{2}\right)\label{eq:B7}
\end{align}
with $L_{1}=L_{2}=-L_{3}=-L_{4}=\left(m-n\right)/2$ for $\left(m-n\right)$
even and $L_{1}=-L_{4}=\left(m-n-1\right)/2$ and $L_{2}=-L_{3}=\left(m-n+1\right)/2$
for $\left(m-n\right)$ odd. 

Similar relations are found for $\nu_{2}$ replacing $\nu_{R}$ by
$-\nu_{R}$. 

The complex constants in \eqref{eq:12-1} read $\varLambda=\left|\varLambda\right|\exp\left(j\lambda\right)$
and $M=\left|M\right|\exp\left(j\mu\right)$ with

\begin{align}
\left|\varLambda\right| & =2^{2\left(\frac{3d+2}{\gamma d}\right)}\sqrt{\left(\frac{2}{\left|d\right|}\omega A\right)^{\left(1-\left(\frac{3d+2}{\gamma d}\right)\right)}}\exp\left(-\pi y\right)\frac{\left|G_{1}\right|}{\left|\nu_{1}\right|}\label{eq:B8}\\
\lambda & =\frac{\pi}{4}\left(1-\left(\frac{3d+2}{\gamma d}\right)\right)-4y\ln\left(2\right)+\arg\left(G_{1}\right)-\arg\left(\nu_{1}\right)\label{eq:B9}\\
\left|M\right| & =2^{-2\left(\frac{3d+2}{\gamma d}\right)}\sqrt{\left(\frac{2}{\left|d\right|}\omega A\right)^{\left(1+\left(\frac{3d+2}{\gamma d}\right)\right)}}\exp\left(-\pi y\right)\frac{\left|G_{2}\right|}{\left|\nu_{2}\right|}\label{eq:B10}\\
\mu & =\frac{\pi}{4}\left(1+\left(\frac{3d+2}{\gamma d}\right)\right)-4y\ln\left(2\right)+\arg\left(G_{2}\right)-\arg\left(\nu_{2}\right)\label{eq:B11}
\end{align}
where $\left|G\right|$ and $\arg\left(G\right)$ are the modulus
and the argument of the complex valued function $G\left(\nu\right)$.

\section*{Appendix C\label{sec:Appendix-C}}

We verify the conditions \eqref{eq:28-1} of application of the WKB
physical optics approximation for the moduli
\begin{equation}
\left|S_{1}\left(R\right)\right|<<\left|\frac{S_{0}\left(R\right)}{\epsilon}\right|\,\,\,;\,\,\left|\epsilon S_{2}\left(R\right)\right|<<\left|S_{1}\left(R\right)\right|\,\,\,;\,\,\left|\epsilon S_{2}\left(R\right)\right|<<1\label{eq:C1}
\end{equation}
with
\begin{align}
\left|\frac{S_{0}\left(R\right)}{\epsilon}\right| & =W\ln\left(2\left(1+\frac{1}{\sqrt{2}}\right)\sqrt{\frac{\omega A}{B}R}\right)+\frac{2}{3}\sqrt{\left(\omega A\right)^{2}R^{3}+B^{2}R-\frac{3}{4}}\label{eq:C2}\\
\left|S_{1}\left(R\right)\right| & =\frac{1}{4}\sqrt{\pi^{2}+\left(\ln\left(\left(\frac{\omega A}{RB^{2}}\right)^{2}\left(\left(\omega A\right)^{2}R^{3}+B^{2}R-\frac{3}{4}\right)\right)\right)^{2}}\label{eq:C3}\\
\left|\epsilon S_{2}\left(R\right)\right| & =\frac{1}{32}\intop^{R}\left(\frac{4\left(-2B^{2}x+\frac{9}{2}\right)\left(\left(\omega A\right)^{2}x^{3}+B^{2}x-\frac{3}{4}\right)}{x\left(\left(\omega A\right)^{2}x^{3}+B^{2}x-\frac{3}{4}\right)^{\frac{5}{2}}}\right.\nonumber \\
 & +\left.\frac{5\left(\left(\omega A\right)^{2}x^{3}+B^{2}x\right)^{2}+15\left(\left(\omega A\right)^{2}x^{3}-B^{2}x+\frac{3}{4}\right)}{x\left(\left(\omega A\right)^{2}x^{3}+B^{2}x-\frac{3}{4}\right)^{\frac{5}{2}}}\right)dx\label{eq:C4}
\end{align}
and $W$ is given by \eqref{eq:50-1}. The first condition \eqref{eq:C1}
reads
\begin{equation}
\left|\frac{S_{0}\left(R\right)}{\epsilon}\right|-\left|S_{1}\left(R\right)\right|>>0\label{eq:C5}
\end{equation}
One easily verifies that the left hand side of the inequality tends
towards positive infinity either when taking the limit for $R\rightarrow+\infty$
with $\epsilon$ constant or when taking the limit for $\epsilon\rightarrow0$
(or $\omega\rightarrow0$) with $R$ constant. The verification of
the second and third conditions implies the solution of the uneasy
elliptic integral \eqref{eq:C4}. One can get some insights into the
verification of these two conditions without solving \eqref{eq:C4},
although, strictly speaking, this method is not exactly rigorous.
Looking at the behaviour of the dominant terms, we take the limit
for $\epsilon\rightarrow0$ (or $\omega\rightarrow0$) under the integral
sign, which yields
\begin{equation}
\left|\epsilon S_{2}\right|\approx\frac{1}{8}\int^{R}\left(\frac{-2B^{2}}{\sqrt{\left(B^{2}x-\frac{3}{4}\right)^{3}}}+\frac{3}{4}\frac{1}{x\sqrt{\left(B^{2}x-\frac{3}{4}\right)^{3}}}+\frac{5}{4}\frac{B^{4}x}{\sqrt{\left(B^{2}x-\frac{3}{4}\right)^{5}}}\right)dx\label{eq:C6}
\end{equation}
which solves easily in
\begin{equation}
\left|\epsilon S_{2}\right|\approx\left|\frac{1}{16}\left(\frac{5}{4}\frac{1}{\sqrt{\left(B^{2}R-\frac{3}{4}\right)^{3}}}+\frac{1}{\sqrt{B^{2}R-\frac{3}{4}}}-\frac{1}{2\sqrt{3}}\arctan\left(\frac{1}{\sqrt{\frac{4}{3}B^{2}R-1}}\right)\right)\right|\label{eq:C7}
\end{equation}
The second condition, for $\epsilon\rightarrow0$ (or $\omega\rightarrow0$),
is always satisfied, provided that $B^{2}R\neq3/4$ for all $R$,
as $\left|S_{1}\right|\rightarrow+\infty$. The third condition is
also satisfied provided that $B^{2}R>>3/4$ for all $R$, which is
the condition \eqref{eq:35-1}. Taking now the limit for $R\rightarrow+\infty$
in \eqref{eq:C7} yields that $\left|\epsilon S_{2}\right|\rightarrow0$,
which satisfies both the second and third conditions \eqref{eq:C1}.


\begin{thebibliography}{1}
\bibitem[1]{Paper1}Pletser V., 2022, ``Annular structures in perturbed
low mass disc-shaped gaseous nebulae I : general and standard models'',
Astrophysics and Space Sciences, submitted.

\bibitem[2]{Pletser1990}Pletser V., 1990, \textquotedbl On exponential
distance relations in planetary and satellite systems, observations
and origin\textquotedbl , PhD Thesis, Physics Dept, Faculty of Sciences,
Catholic University of Louvain, Louvain-la-Neuve, Belgium (available
at https://www.researchgate.net/publication/257927392).

\bibitem[3]{Whittaker-Watson1927}Whittaker, E.T. \& Watson, G.N.,1927,
A course of Modern Analysis , 4th ed., Cambridge, 337.

\bibitem[4]{Kamke1943}Kamke, E., 1943, Differentialgleichungen, aufl.2,
Akad. Verlagsges. Seeker and Erler Kom. -ges., Leipzig, 473. 

\bibitem[5]{Jahnke-Emde-Losch1966}Jahnke-Emde-Losch, 1966, ``Tafeln
h6herer Funktionen'', B.G. Teubner Verlagsges, Stuttgart. 

\bibitem[6]{Slater1965}Slater, L.J., 1965, in ``Handbook of Mathematical
Functions:, eds M. Abramowitz and I. Stegun, Dover Publ., New York,
503. 

\bibitem[7]{Olver1965}Olver, F.W.J., 1965, in ``Handbook of Mathematical
Functions'', eds M. Abramowitz and I. Stegun, Dover Publ., New York,
355. 

\bibitem[8]{Bender-Orszag1978}Bender, C.M., \& Orszag, S.A., 1978,
Advanced mathematical methods for Scientists and Engineers , McGraw-
Hill Book Co., 484. 

\bibitem[9]{Gradshteyn-Ryzhik1965}Gradshteyn, I.S., \& Ryzhik, I.M.,
1965, Tables of Integrals, Series, and Products, transl. from Russian
by A.Jeffrey (ed.), Academic Press, New York, 217. 
\end{thebibliography}
\end{document}